\documentclass[conference]{IEEEtran}
\IEEEoverridecommandlockouts
\usepackage{comment}
\usepackage{cite}
\usepackage{url}
\usepackage{amsmath,amssymb,amsfonts}
\usepackage{algorithmic}
\usepackage{diagbox}
\usepackage{graphicx}
\usepackage{textcomp}
\usepackage{multirow}
\usepackage{adjustbox}
\usepackage{xcolor}
\usepackage{subfigure}
\usepackage{hyperref} 
\usepackage{float} 

\def\BibTeX{{\rm B\kern-.05em{\sc i\kern-.025em b}\kern-.08em
    T\kern-.1667em\lower.7ex\hbox{E}\kern-.125emX}}

\begin{document}

\title{FG-DFPN: Flow Guided Deformable Frame Prediction Network\\
\thanks{A. M. Tekalp acknowledges support from the Turkish Academy of Sciences (TUBA).}
}

\author{\IEEEauthorblockN{M. Akın Yılmaz}
\IEEEauthorblockA{\textit{Codeway AI Research} \\
Istanbul, Turkey \\
mustafaakinyilmaz@ku.edu.tr}
\and
\IEEEauthorblockN{Ahmet Bilican}
\IEEEauthorblockA{\textit{Electrical \& Electronics Eng.}\\ \textit{Koç University}\\ Istanbul, Turkey\\
abilican21@ku.edu.tr}
\and
\IEEEauthorblockN{A. Murat Tekalp}
\IEEEauthorblockA{\textit{Electrical \& Electronics Eng.}\\ \textit{Koç University}\\ Istanbul, Turkey\\
mtekalp@ku.edu.tr}
}

\maketitle

\begin{abstract}
Video frame prediction remains a fundamental challenge in computer vision with direct implications for autonomous systems, video compression, and media synthesis. We present FG-DFPN, a novel architecture that harnesses the synergy between optical flow estimation and deformable convolutions to model complex spatio-temporal dynamics. By guiding deformable sampling with motion cues, our approach addresses the limitations of fixed-kernel networks when handling diverse motion patterns. The multi-scale design enables FG-DFPN to simultaneously capture global scene transformations and local object movements with remarkable precision. Our experiments demonstrate that FG-DFPN achieves state-of-the-art performance on eight diverse MPEG test sequences, outperforming existing methods by 1dB PSNR while maintaining competitive inference speeds. The integration of motion cues with adaptive geometric transformations makes FG-DFPN a promising solution for next-generation video processing systems that require high-fidelity temporal predictions. The model and instructions to reproduce our results will be released at: \normalfont{\url{https://github.com/KUIS-AI-Tekalp-Research-Group/frame-prediction}}.
\end{abstract}


\section{Introduction}
\label{intro}
Video frame prediction aims to forecast future video frames based on one or more past frames. While traditionally formulated as a supervised learning task, it inherently benefits from self-supervision since the ground-truth future frames are available within the video sequence. In complex scenarios—where both fast and slow moving objects, camera motion, and abrupt scene changes coexist—capturing intricate motion patterns becomes particularly challenging.

Our proposed Flow-Guided Deformable Frame Prediction Network (FG-DFPN) harnesses the power of optical flow to guide deformable convolution operations. In our design, a dedicated flow estimator module produces coarse motion cues that serve as a reference for learning adaptive offset maps. These offsets dynamically adjust the sampling locations of the convolutional kernels, allowing the network to effectively model non-rigid and multi-scale motion. This flow-guided approach not only stabilizes training but also significantly improves the alignment of multi-scale features, leading to sharper and more accurate future frame predictions.

Applications of frame prediction span autonomous driving, where real-time forecasting is critical, to video compression, where precise frame synthesis can reduce the need for transmitting explicit motion information. As demonstrated in Table ~\ref{table:results}, FG-DFPN achieves significantly higher PSNR values compared to all previous video prediction networks, particularly preserving fine details in regions with complex motion and being much faster than all previous methods, which is critical in most contexts.

\begin{figure*}[t]
\centering
	\includegraphics[width=0.96\textwidth]{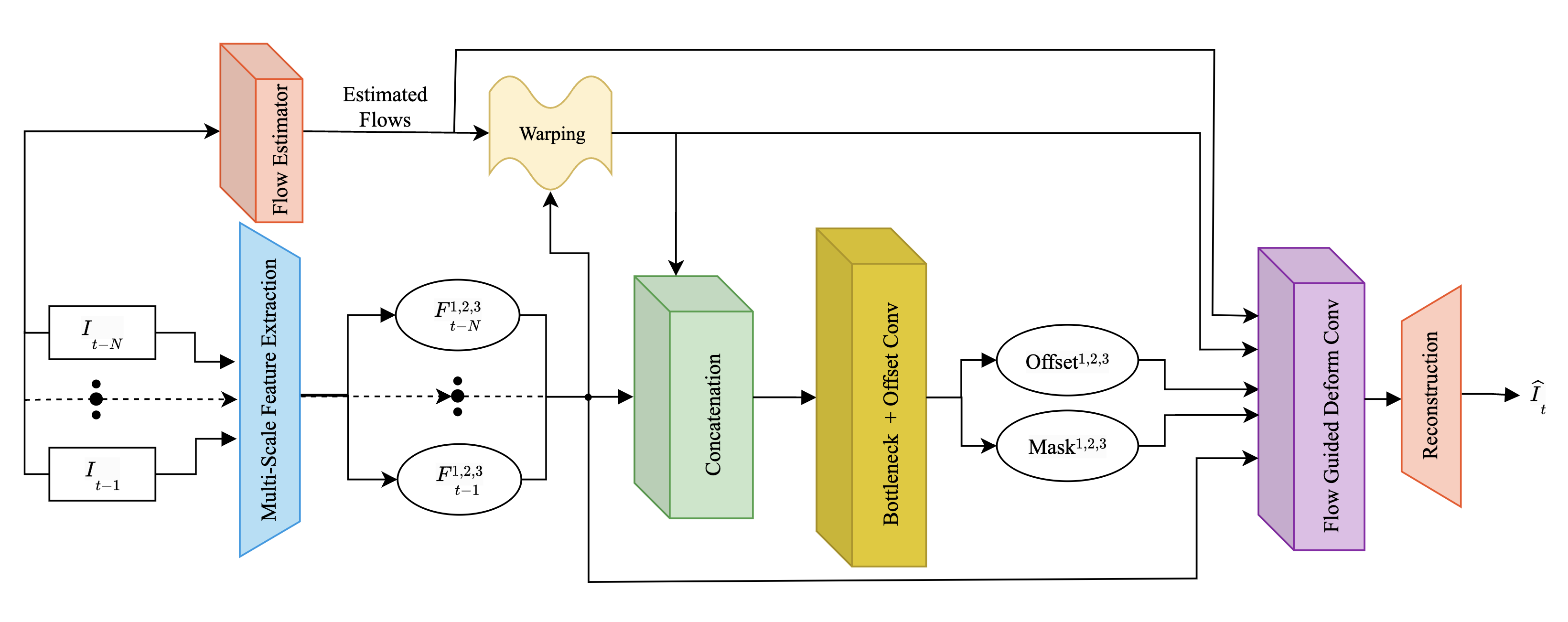} \vspace{-2pt} \\

\caption{Our proposed FG-DFPN framework}
\label{fig:our_model}
\end{figure*}

\section{Related work and Contributions}
\label{related}
\subsection{\textbf{Frame Prediction}} \vspace{-2pt}
Video frame prediction has been explored through various paradigms, as summarized in comprehensive reviews~\cite{review1, review2}. Traditional approaches often rely on recurrent convolutional encoder-decoder architectures~\cite{shi, fcnn} or 3D convolutions~\cite{crevnet} to capture temporal dynamics. Some methods formulate frame prediction as a direct synthesis problem in the pixel domain~\cite{shi, fcnn, serkan, diffcnn}, while others explicitly model inter-frame transformations~\cite{fstn, sdcnet}.

Recent works have further diversified the landscape by leveraging transformation-based and hybrid architectures as well as lightweight designs. Mokssit et al.\cite{Mokssit2025} propose a spatial transformation-based next frame predictor that disentangles object motion from camera motion by estimating per-object affine and projective transforms—requiring only 14 parameters per object—to robustly predict future frames even under occlusions. Mathai et al.\cite{Mathai2024Hybrid} introduce 3DTransLSTM, a hybrid transformer-LSTM model that employs 3D separable convolutions to efficiently capture long-range spatiotemporal dependencies while significantly reducing model size and computational cost. In a related vein, Mokssit et al.\cite{Mokssit2024ObjectOriented} develop an object-oriented framework that estimates transformation parameters for each detected object, operating in a low-dimensional transformation space rather than directly in the pixel domain, thereby enhancing prediction robustness in dynamic scenes. Complementing these efforts, Mathai et al.\cite{Mathai2022Lightweight} present a lightweight model that integrates separable CNNs with an ST-LSTM architecture to achieve competitive accuracy with a greatly reduced parameter count and complexity. Extending the scope beyond traditional video prediction, Hudson et al.~\cite{Hudson2024EverythingVideo} reformulate a variety of multimodal tasks—spanning text, images, audio, and video—as a unified next-frame prediction problem, enabling a single transformer-based model to seamlessly integrate and process different modalities.

Despite progress, many existing approaches still struggle with accurately modeling complex motion, and losses based solely on $l_{1}$ or $l_{2}$ norms~\cite{fcnn, sdcnet} may lead to blurry predictions. This has spurred research into perceptual and adversarial losses to improve visual fidelity~\cite{sdcnet, villegas}. The paper~\cite{dfpn} uses deformable convolutions for modeling geometric transformations, demonstrating that such an approach achieves good performance; we build upon this finding in this paper by enhancing training resilience and improving performance by over 1dB.
\subsection{\textbf{Flow-Guided Deformable Convolutions}} \vspace{-2pt}
Deformable convolution~\cite{deform_cnn} introduces learnable offsets to the conventional convolutional grid, allowing the network to adapt its receptive field to varying motion patterns. Deformable convolution applied to object detection and video restoration~\cite{tdan, edvr}, video compression \cite{icip23}, and video prediction \cite{dfpn}. Using flow guidance for the offsets have been demonstrated to achieve much better results in video super-resolution \cite{chan2022basicvsrpp}, and compression \cite{icip2024}. In FG-DFPN, optical flow is used to guide the deformable convolutions: a Flow Estimator (FlowNET) predicts coarse flow, which is then incorporated into offset prediction for the deformable convolution layers. This strategy ensures that the learned offsets remain consistent with the underlying motion, effectively aligning features across frames and mitigating misalignment artifacts inherent in pure flow-based warping.

\begin{figure*}[!t]
    \centering
    \subfigure[]{\includegraphics[width=0.28\textwidth]{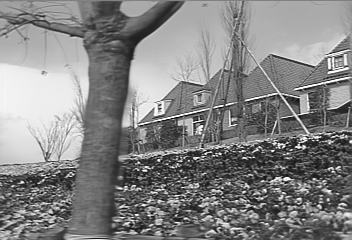}}  
    \subfigure[]{\includegraphics[width=0.28\textwidth]{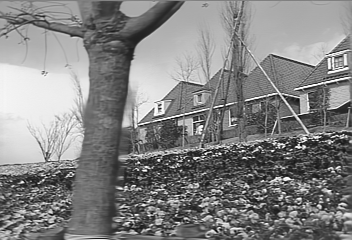}}   
    \subfigure[]{\includegraphics[width=0.28\textwidth]{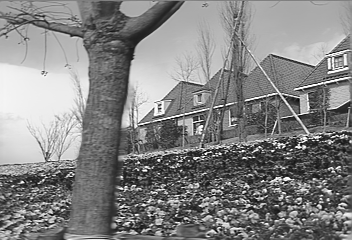}} 
    \subfigure[]{\includegraphics[width=0.28\textwidth]{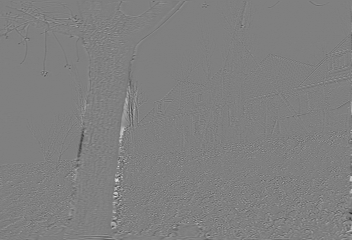}} 
    \subfigure[]
    {\includegraphics[width=0.28\textwidth]{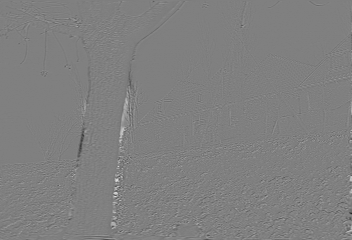}} 
    \subfigure[]{\includegraphics[width=0.28\textwidth]{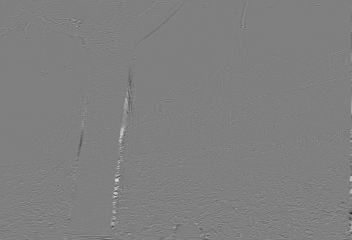}} 
\vspace{-2pt}
    \caption{Visual comparison of predicted frames for the \textit{Garden} sequence. Top row shows the prediction results from: (a) DiffCNN (PSNR: 26.48, SSIM: 0.922), (b) DFPN (PSNR: 26.76, SSIM: 0.936), and (c) our proposed FG-DFPN (PSNR: 28.78, SSIM: 0.964). Bottom row (d-f) displays the corresponding error maps between the ground truth and predictions from each method, clearly demonstrating the superior performance of our approach with significantly reduced errors.} \vspace{-10pt}
    \label{fig:image64}
\end{figure*}

\subsection{\textbf{Contributions}} \vspace{-2pt}
Our main contributions are summarized as follows:
\begin{itemize}
    \item We propose the first Flow-Guided Deformable Frame Prediction Network (FG-DFPN) that combines explicit optical flow guidance with adaptive deformable convolutions to capture complex spatiotemporal dynamics.
    \item We introduce a multi-scale architecture that integrates a Flow Estimator, a Multi-Scale Feature Extractor, an Offset and Mask Predictor, and novel Offset Diversity modules. This design allows the network to fuse both global motion trends and fine local details, leading to superior prediction performance.
    \item Extensive experiments demonstrate that FG-DFPN outperforms conventional fully convolutional~\cite{fcnn, diffcnn} and deformable transformation based\cite{dfpn} networks in terms of prediction accuracy, runtime ,and computational efficiency, making it a promising solution for applications such as autonomous driving and video compression.
\end{itemize}

\section{Proposed Method}
\label{method}

Our video prediction framework processes four consecutive grayscale frames and is built from several key components that work in tandem to capture multi-scale motion and fuse information via deformable convolutions. Figure~\ref{fig:our_model} provides an overview of the architecture. In brief, the system first estimates coarse motion using an optical flow estimator, then extracts multi-scale features from each input frame, warps these features according to the estimated flows, predicts deformable offsets and modulation masks, fuses the aligned features via flow-guided deformable convolutions, and finally reconstructs the predicted frame.

\subsection{Flow Estimation}
A dedicated flow estimation module receives the four input frames (stacked channel-wise) and computes coarse optical flow fields. The network is designed with several downsampling and upsampling stages, employing residual bottleneck blocks, to yield a flow prediction for each input frame. These flows serve as initial motion cues, which are later used both for warping features and for guiding the refinement of spatial offsets.

\subsection{Multi-Scale Feature Extraction and Warping}
Each input frame is independently processed by a multi-scale feature extractor that outputs three levels of feature maps (denoted as \(f^l\) for \(l=1,2,3\)). The first level retains high spatial resolution, while levels two and three are obtained via successive strided convolutions to capture coarser context. For each scale, the corresponding optical flow is adapted via bilinear interpolation (to match the resolution of the feature maps) and is used to warp the extracted features. Importantly, by performing warping in the feature space rather than in the pixel space, our approach avoids the interpolation artifacts typically observed when warping is applied directly to pixel values. The warping function generates a warped version \(w^l\) of the original feature map \(f^l\) by constructing a sampling grid that is shifted by the flow displacement, yielding both the original and warped features at each scale.

\subsection{Offset and Mask Prediction}
To guide the deformable feature fusion, the network estimates spatial offsets that capture fine-grained motion. For each scale, the original features \(f^l\) and their warped counterparts \(w^l\) from all four frames are concatenated to form a composite feature tensor. This tensor is then processed through a series of downsampling layers, bottleneck residual blocks, and upsampling layers that output, for each scale, a set of offset predictions and corresponding modulation masks. The offsets are predicted separately for each input frame and are initialized such that the initial deformation is zero.

\begin{table*}[!t]
\centering
\caption{\textbf{Quantitative comparison (PSNR/SSIM).} {\color{red}Red} and {\color{blue}blue} colors indicate the best and the second-best performance, respectively} \vspace{-5pt}
\begin{adjustbox}{width=0.9\textwidth}
\begin{tabular}{l|c|c|c|c|c}
\hline
\multicolumn{1}{c|}{\textbf{}} &
  \begin{tabular}[c]{@{}c@{}}FCNN\\ \cite{serkan}\end{tabular} &
  \begin{tabular}[c]{@{}c@{}}DiffCNN\\ \cite{diffcnn}\end{tabular} &
  \begin{tabular}[c]{@{}c@{}}CLSTM\\ \cite{fcnn}\end{tabular} &
  \begin{tabular}[c]{@{}c@{}}DFPN\\ \cite{dfpn}\end{tabular} &
  \textbf{\begin{tabular}[c]{@{}c@{}}FG-DFPN\\ (Ours)\end{tabular}} \\ \hline
Coastguard       & 32.00 / 0.915 & \color{blue}32.08 / 0.917 & 30.55 / 0.885 & 32.00 / 0.914 & \color{red}32.75 / 0.934 \\ \hline
Container        & 41.50 / 0.982 & \color{red}42.08 / 0.983 & 40.77 / 0.981 & 41.78 / 0.981 & \color{blue}41.99 / 0.983 \\ \hline
Football         & {\color{blue}22.87} / 0.776 & 22.71 / 0.778 & 22.01 / 0.729 & 22.72 / {\color{blue}0.779} & \color{red}23.26 / 0.804 \\ \hline
Foreman          & {\color{blue}31.38} / 0.890 & 31.29 / 0.893 & 30.06 / 0.857 & 31.26 / {\color{blue}0.897} & \color{red}33.07 / 0.935 \\ \hline
Garden           & 26.45 / 0.931 & 26.42 / 0.928 & 22.54 / 0.831 & \color{blue}27.30 / 0.940 & \color{red}29.15 / 0.963 \\ \hline
Hall Monitor     & {\color{red}36.60} / {\color{blue}0.961} & {\color{blue}36.53} / 0.959 & 35.65 / 0.955 & 36.24 / 0.954 & 36.44 / {\color{red}0.963} \\ \hline
Mobile           & 27.31 / 0.949 & 27.79 / 0.956 & 23.99 / 0.884 & \color{blue}28.16 / 0.961 & \color{red}28.79 / 0.969 \\ \hline
Tennis           & 29.88 / 0.872 & 29.90 / 0.882 & 28.27 / 0.826 & \color{blue}30.06 / 0.886 & \color{red}31.24 / 0.913 \\ \hline \hline 
\textbf{Average} & \textbf{31.00 / 0.909} & \textbf{31.10 / 0.912} & \textbf{29.23 / 0.869} & \textbf{\color{blue}31.19 / 0.914} & \textbf{\color{red}32.09 / 0.933} \\ \hline \hline 
Params (M)       & 38.4     & 38.4 & 1.0 & 5.8 & 45.8 \\ \hline
Runtime (ms)     & 950     & 950 & 40 & 165 & 125  \\ \hline
\end{tabular}
\end{adjustbox}
\label{table:results}
\end{table*}

\subsection{Flow-Guided Offset Refinement and Deformable Fusion}
Given the predicted offsets and the corresponding flows at each scale, the model refines the deformation parameters via a flow-guided offset refinement process. For each scale \(l\), the procedure is as follows:
\begin{itemize}
    \item The predicted output is split into two offset components and a modulation mask.
    \item A hyperbolic tangent nonlinearity, scaled by a magnitude parameter, is applied to constrain the offset range.
    \item The offsets are adjusted by adding the corresponding flow (after appropriate channel-wise processing) to ensure consistency with the estimated motion.
\end{itemize}
Once refined, the offsets and masks from the four frames are concatenated and used in a deformable convolution operation (implemented via a grouped deformable convolution layer) that fuses the individual feature maps \(f^l\) into a single, refined feature representation \(\hat{f}^l\) at each scale. Unlike pixel-space warping—which can produce artifacts due to interpolation—the deformable convolution not only avoids these issues but also models complex geometric transformations more effectively by learning adaptive spatial offsets. The use of different magnitude parameters (e.g., 10, 20, and 40 for the coarsest to finest scales) allows the network to flexibly handle motions of varying extent.

\subsection{Reconstruction}
Finally, the refined multi-scale features \(\hat{f}^1\), \(\hat{f}^2\), and \(\hat{f}^3\) are input to the reconstruction module. This module employs additional residual bottleneck blocks and subpixel convolution layers to progressively fuse and upsample the features, ultimately synthesizing the predicted next frame. The output is a single-channel image that represents the forecasted frame in the video sequence.

\section{Experiments}
\label{experiments}

\subsection{Experimental Setup}
\subsubsection{\textbf{Training Settings}}
We train our deformable frame prediction network on gray-scale frames extracted from Vimeo-90k septuplets\cite{vimeo}. We randomly select 5 consecutive frames and apply $96 \times 96$ random cropping to the same positions in order to create an augmented batch of consecutive patches. The first 4 frames of a batch are used as input to our model and the 5th frame is taken as the ground-truth frame. We use Charbonnier loss\cite{charbonnier} since it better handles outliers and improves the performance over the conventional $L_{2}$ loss. We set the batch size as 8 and used Adam\cite{adam} optimizer for 500K iterations. Initial learning rate is set to 0.0001 and halved in every 100K iterations.

\subsubsection{\textbf{Test Sequences and Evaluation}}
We test the performance of our model on 8 MPEG sequences: Coastguard, Container, Football, Foreman, Garden, Hall Monitor, Mobile, and Tennis and compare the results with those in \cite{serkan, fcnn, diffcnn, dfpn}. Per-sequence and average PSNR\cite{psnr_comp} and SSIM\cite{ssim} are reported for quantitative results.

\subsection{Experimental Results} \label{exp_results}
Table~\ref{table:results} reports a quantitative comparison of five frame prediction methods, including our proposed FG-DFPN, on eight MPEG test sequences. We present both PSNR (in dB) and SSIM values, with the best and second-best scores highlighted in \textcolor{red}{red} and \textcolor{blue}{blue}, respectively.

\subsubsection{Quantitative Performance} FG-DFPN attains the highest overall results, achieving an average PSNR/SSIM of \textbf{32.09 dB / 0.933}. This represents a clear improvement over all previous methods. In particular, notable gains are observed on challenging sequences such as \emph{Garden} and \emph{Tennis}, where the proposed approach outperforms the previous best results by large margins in PSNR and SSIM. While DiffCNN excels in the \emph{Container} sequence with 42.08 dB, FG-DFPN remains highly competitive at 41.99 dB. Similarly, for \emph{Hall Monitor}, FG-DFPN trades off a slight PSNR difference (36.44 dB vs. 36.60 dB for FCNN) for better SSIM (0.963 vs. 0.961), underscoring its balanced performance across different sequences.

\subsubsection{Model Complexity and Runtime} Though FG-DFPN has the largest number of parameters (45.8 M) among the compared methods, its carefully optimized deformable architecture enables a fast runtime of only 125 ms per frame. This is significantly lower than the 950 ms required by the similarly sized FCNN and DiffCNN, highlighting that raw parameter counts do not solely determine speed. Instead, well-designed network components contribute to lower latency. Although CLSTM has the smallest parameter count (1.0 M) and fastest runtime (40 ms), its performance in terms of average PSNR/SSIM (29.23 dB / 0.869) lags behind FG-DFPN’s state-of-the-art results.

Overall, the effectiveness of FG-DFPN can be attributed not just to the number of parameters but also to the design of its deformable modules and feature fusion mechanisms. These allow the model to capture complex motion dynamics while maintaining a favorable trade-off between accuracy and speed.

\vspace{10pt}



\section{Conclusion}
\label{conclusions}

In this work, we proposed FG-DFPN, a novel flow-guided deformable framework for video frame prediction that effectively captures temporal correlations. Our extensive evaluation on eight MPEG test sequences demonstrates that FG-DFPN achieves superior performance in terms of PSNR and SSIM compared to existing methods while maintaining competitive inference speed. In particular, despite having the largest number of parameters among the models compared, the carefully optimized architecture design of FG-DFPN facilitates significantly faster runtimes than some heavier counterparts. These results highlight the importance of both network capacity and design efficiency in achieving high-quality video frame predictions under practical time constraints. Future research will explore more advanced fusion strategies and lightweight modules to further enhance both prediction accuracy and efficiency.
\clearpage
\bibliographystyle{IEEEtran}
\bibliography{references}
\end{document}